\documentclass[12pt]{article}
\usepackage{amssymb}
\usepackage{a4wide}

\def\be{\begin{equation}}
\def\ee{\end{equation}}

\def\RSa{1}
\def\RSb{2}
\def\PK{3}
\def\DI{4}
\def\DD{5}
\def\MSM{6}
\def\NOZ{7}
\def\KS{8}
\def\GS{9}
\def\CH{10}
\def\Is{11}
\def\ACO{12}

\begin{document}

\title{Colliding Bubble Worlds}

\author{
Warren B. Perkins\footnote{w.perkins@swansea.ac.uk}  \\ \\
\em Department of Physics, \\
\em University of Wales Swansea,\\
\em Singleton Park, Swansea, SA2 8PP, \\
\em Wales
}

\date{\today}
\maketitle
\vskip -3.5truein
\hfill{SWAT/268}
\vskip 3.5truein
\pagenumbering{arabic}
\begin{abstract}
We consider a cosmological model in which our Universe is a spherically
symmetric bubble wall in 5-dimensional anti-de Sitter spacetime. We argue
that the bubble on which we live will undergo collisions with other similar
bubbles and estimate the spectrum of such collisions. The collision rate is found to 
be independent of the age of our Universe. Collisions with small bubbles
provide an experimental signature of this scenario, while collisions with larger
bubbles would be catastrophic. 
\end{abstract}

\section{Introduction}

There has been a great deal of interest recently in higher dimension
theories with non-compact extra dimensions [\RSa,\RSb]. Rather 
than compactify the extra dimensions to leave the three  observed
large spatial dimensions, matter is localised to a 3-brane. While
gravity exists in the bulk, there is a gravitational zero-mode on
the brane which mimics four dimensional gravity at low energy. Our
low energy world thus appears to be four dimensional.

While many particle physics constraints on such models have been
addressed in the context of  {\it plane}  branes [\RSa,\RSb], it is also possible 
for these branes to be closed, expanding domain walls [\PK,\DI,\DD]. These two
pictures are closely related [\MSM]. Both plane branes and bubbles
can be created, complete with their  5-dimensional anti-de Sitter
(AdS$_5$) bulk spacetimes, from nothing, by the 
appropriate instantons [\NOZ,\KS,\GS]. 

In this paper we consider an alternative cosmological view in which our
brane is an expanding bubble nucleated in a pre-existing AdS$_5$ bulk.
In the spirit of more conventional bubble nucleation settings, we consider
the bubble to separate two regions of AdS$_5$ with differing cosmological
constants. The effective Friedmann equation on our brane is considered in
section 2. In this scenario,  our bubble is simply one of many bubbles 
forming in the AdS$_5$ bulk. In section 3 we consider the spectrum of 
bubble-bubble collisions. While providing possible experimental evidence
for our scenario, there is also the possibility of a catastrophic
bubble-bubble collision in our neighbourhood.

\section{Effective Friedmann Equation On The Bubble}   

We consider a bubble separating two AdS$_5$ regions with different cosmological
constants, $\Lambda_\pm$. The effective Friedmann equation in this situation has 
been analysed in ref.\DD, while the linearized gravity on such branes is discuused
in ref.\CH. Here, in order to specify our model, we review the results of ref.\DD.

The metrics on either side of the bubble have the form,
\be
ds^2=-(k+H_\pm^2A^2)L_\pm(t)dT^2 +(k+H_\pm^2A^2)^{-1}dA^2 
+A^2[d\chi^2+ f_k^2(\chi)(d\theta^2+\sin^2(\theta)d\phi^2)],
\label{metric}
\ee
where,
\be
H_\pm^2=-{\Lambda_\pm\over 6},
\ee
$f_{-1}(\chi)=\sinh{\chi}$, $f_{0}(\chi)=1$ and $f_{1}(\chi)=\sin{\chi}$.
$L_\pm(t)$ are lapse functions that 
allow the coordinates to be matched at the bubble. For simplicity we set $L_+(t)=1$. 

If we let the position of the bubble be $T=t(\tau)$,
$A=a(\tau)$, where $\tau$ is the proper time on the bubble, the induced
metric on the bubble is,
\be
ds_4^2=-d\tau^2 +a(\tau)^2[d\chi^2+ f_k^2(\chi)(d\theta^2+\sin^2(\theta)d\phi^2)].
\ee
As the  bulk cosmological constants differ on either side of the bubble, the
Israel matching condition [\Is] is used to relate the change in the extrinsic
curvature at the bubble to the energy-momentum of the bubble. Assuming a
perfect fluid form for the stress-energy tensor on the bubble,
$T^\mu_\nu={\rm diag}(-\rho,p,p,p)$, the effective Friedmann equation on
the bubble is found to be[\DD],
\be
{8\over 3}\pi G\rho a= \sqrt{{\dot a ^2}+k-{\Lambda_+\over 6}a^2}
                      -\sqrt{{\dot a ^2}+k-{\Lambda_-\over 6}a^2}.
\ee
This can be cast into a more familiar form by splitting
the bubble stress-energy into a constant tension part and a
matter contribution,
\be
\rho=\rho_m +\sigma.
\ee

At late times,  $\rho_m<<\sigma$ and it is possible to  expand the right hand side of
the effective Friedmann equation as a power series in $\rho$, 
\be
H^2+{k\over R^2}= \Lambda_{\rm eff.}+ \Phi \rho_m +\Psi \rho_m^2 + ...  
\label{Fried}
\ee
The effective cosmological constant can be set to zero by tuning the bubble tension, 
$\sigma$, according to the constraint,
\be
\bigl({8\over 3} \pi G\sigma\bigr)^2=\bigl[\sqrt{-\Lambda_+\over 6}\pm
                                           \sqrt{-\Lambda_-\over 6}\bigr]^2.
\ee
The coefficients $\Phi$ and $\Psi$ are then given by, 
\be
\Phi={4\over 3}\pi G
\biggl[
4 {\sqrt{{\Lambda_+\over 6}{\Lambda_-\over 6}}\over
      \vert \sqrt{-\Lambda_+\over 6}\pm\sqrt{-\Lambda_-\over 6} \vert} 
\biggr],
\quad
\Psi={16\over 9}\pi^2 G^2 \biggl[ 1+3
{(\sqrt{-\Lambda_+} \mp\sqrt{-\Lambda_-})^2 \over
[\sqrt{-\Lambda_+}\pm\sqrt{-\Lambda_-}]^2}
\biggr],
\ee
where the overall sign ambiguity in taking the square root of $\sigma^2$ has been resolved
by requiring that both $\sigma$ and $\Phi$ are positive. There remain two possible
values for the bubble tension,
\be
\sigma_+={3\over 8\pi G}[\sqrt{-\Lambda_+\over 6}+\sqrt{-\Lambda_-\over 6}],
\quad
\sigma_-={3\over 8\pi G}\vert\sqrt{-\Lambda_+\over 6}-\sqrt{-\Lambda_-\over 6}\vert.
\ee
The latter solution is discussed in ref. \DD.

In either case, at late times, we have a Friedmannian evolution of the scale factor 
on our bubble. As usual in these scenarios, there is also a non-zero coefficient of 
$\rho_m^2$ and the effective four dimensional Planck 
scale is determined by the fundamental Planck scale and the bulk cosmological 
constants.

\section{Bubble-Bubble collisions}

If we live on an expanding bubble that was nucleated in AdS$_5$, we must consider
the possibility of other bubbles forming in the same manner and colliding with ours. 
The nucleation process in the special case, $\Lambda_+ =\Lambda_-$, has been 
considered [\ACO]. However, without a concrete underlying model, we cannot determine
the tunnelling probability or bubble nucleation rate, but we can estimate the
spectrum of bubble-bubble collisions.

We assume that the nucleation rate per unit volume, $\Gamma$, is constant, so that the 
probability of a nucleation event in the region $T\to T+dT$, $A\to A+dA$, 
$\chi\to\chi +d\chi$, $\theta\to\theta+d\theta$, $\phi\to\phi+d\phi$ is
$\Gamma A^3f^2_k(\chi)\sin{(\theta)}dT dA d\chi d\theta d\phi$. If we only consider
small bubbles colliding with a large bubble, we are only interested in nucleation events in
the vacinity of the large bubble. These have nucleation probability per unit time per unit bubble
volume of approximately $\Gamma$.
 
The evolution of a single bubble in the AdS$_5$ has been determined in ref.\DD. 
In terms of the bubble proper time $\tau$, the position of the shell is given by,
\be
A=a(\tau), \quad T=t(\tau),
\label{ATtau}
\ee
where $a(\tau)$ is determined by the modified Friedmann equation and,
\be
{dt\over d\tau}={\sqrt{k+H^2a^2 +({da\over d\tau})^2}\over k+H^2a^2}.
\label{taut}
\ee
For simplicity we consider the $k=1$ case with the standard power law approximations 
for the evolution of the scale factor,
\be
a=\alpha \tau^p, \quad p={1\over 4}, {1\over 2}, {2\over 3}.
\ee
The last two values of $p$ correspond to the standard radiation and 
matter dominated eras respectively. The first value holds at early times when the
bubble is radiation dominated and the $\rho_m^2$ term in the effective Friedmann
equation is dominant. 
Using (\ref{taut}), the background metric time at the bubble can then be related to 
the scale factor.
At early times, $({da\over d\tau})^2>>1+H^2a^2$, leading to,
\be
t-t_0 \simeq a,
\ee
while at late times, $H^2a^2>>1+({da\over d\tau})^2$, we find,
\be
t-t_0\sim {1\over H\alpha} {\tau^{1-p}\over 1-p}
\to
a\sim \alpha[(1-p)H\alpha(t-t_0)]^{p\over 1-p}
=\cases{ {H\alpha^2\over 2} (t-t_0) & $p={1\over 2}$ \cr
         {H^2\alpha^3\over 9} (t-t_0)^2 & $p={2\over 3}$\cr}
\ee
While the coordinate spped of the bubble seems to grow at late times,
the physical speed of the bubble wall relative to the bubble centre is close to unity
at early times but drops like $(t-t_0)^{-3}$ at late times.

In order to determine the evolution of a pair of bubbles, we take two such bubbles
with different centres. The centre of the second 
bubble is at the origin of a shifted coordinate system defined by,
$$
\tan{H\tilde T}=\cosh{p}\tan{HT}-\sinh{p}{cos{\chi}\over\cos{HT}}
                                   {AH\over\sqrt{1+A^2H^2}},
$$
$$
\tilde A^2
=A^2\sin^2{\chi}+\bigl[\sinh{p}\sin{HT}\sqrt{{1\over H^2} +A^2}
          -A\cosh{p}\cos{\chi}\bigr]^2,
$$
$$
\cos{\tilde\chi}={-\sinh{p}\sin{HT}\sqrt{{1\over H^2} +A^2} +\cosh{p}A\cos{\chi}
                     \over
                  \bigl[
 A^2\sin^2{\chi}+\bigl(\sinh{p}\sin{HT}\sqrt{{1\over H^2} +A^2}
          -A\cosh{p}\cos{\chi}\bigr)^2          
                  \bigr]^{1/2}          },
$$
\be
\tilde\theta=\theta, \quad \tilde\phi=\phi.
\ee
The metric in these coordinates is precisely that given in (\ref{metric}). 
The origin of this system 
lies on the $\sin{\chi}=0$ line, while the origin of the original system lies on the
 $\sin{\tilde\chi}=0$ line. The initial point of contact of the
two bubble walls also lies on these lines, so we can 
work with the simplified transformations,
$$                     
\tan{H\tilde T}=\cosh{p}\tan{HT}-{\sinh{p}\over\cos{HT}}
                                   {AH\over\sqrt{1+A^2H^2}},
$$
\be
\tilde A=\sinh{p}\sin{HT}\sqrt{{1\over H^2} +A^2}
          -\cosh{p}A.
\label{trans}
\ee       

If a bubble  nucleates at $\tilde A=0$, we can use (\ref {ATtau}) and
(\ref{Fried}) to determine its
evolution in the $(\tilde T,\tilde A,\tilde \chi,\tilde\theta ,\tilde \phi)$ coordinates.
We assume that the centre of this bubble is at rest with respect to the centre
of the first bubble at nucleation.  If this nucleation event occurs at
$T=T_2$, $\tilde T=\tilde T_2$, we have, $HT_2=H\tilde T_2=\pi/2$ and the spacing of the 
centres is given by,
\be
H\tilde A_0=\sinh{p}.
\label{space}
\ee

We consider a case relevant to our current Universe: one old bubble ($p={2/3}$)
colliding with a younger bubble ($p=1/2$). The old bubble has its origin at $\tilde A=0$
and it is large, so the second bubble must nucleate a long away from  $\tilde A=0$. Thus, 
$H\tilde A_0$ is large and from (\ref{space}) we have $\sinh{p}\sim\cosh{p}$.
For simplicity we assume that the time between the nucleation of the second bubble and
the collision of the two bubbles is short compared to $H^{-1}$. Denoting the
elapsed times by $\Delta T$ and $\Delta \tilde T$, we can use (\ref{trans}) to relate the two at the 
position of the second bubble.

To leading order we find,
\be
(H\Delta \tilde T)^2= {(H\Delta T)^2 \over
                  \cosh^2{p}\bigl[\sin{HT}-{AH\over\sqrt{1+A^2H^2}}\bigr]^2 },
\ee
where $A$ and $T$ are the coordinates of the bubble nucleated at $A=0$.
The bubble spacing is given by,
\be
\Delta \tilde A \sim \tilde A_0[\sin{HT}\sqrt{1+A^2H^2} -AH] 
                  - {H^2\alpha^3\over 9}(\tilde T_2 -\tilde T_1)^2,
\ee
where the first term represents the position of the young bubble wall and the 
second is the radius of the old bubble. Assuming that $AH<<1$ and using 
$A\sim H\alpha^2\Delta T/2$ for the young bubble, we find to leading order in 
$\Delta T$ and $\Delta\tilde T$,
$$
\Delta \tilde A \sim \tilde A_0[1-{1\over 2}H^2\alpha^2\Delta T] 
- {H^2\alpha^3\over 9}
\bigl[(\tilde T -\tilde T_1)+2\Delta \tilde T(\tilde T_2 -\tilde T_1)\bigr]
$$
\be
\sim\Delta \tilde A_0 -\Delta \tilde T\biggl\{{\cosh{p}\over 2}
\tilde A_0 H^2\alpha^2
+2{H^2\alpha^3\over 9}(\tilde T_2 -\tilde T_1) \biggr\}.
\ee              
The elapsed time before collision is then,
\be
\delta\tilde T \propto {\Delta \tilde A_0\over  \tilde A_0^2}.
\ee
In the case of young bubbles colliding with an old one, $\tilde A_0$ is  
roughly the scale factor of the old bubble, so we take it to be
constant for all young bubbles hitting at an instant.  

We denote the number of bubbles impacting with radius between $\tilde R_{im}$ and 
$\tilde R_{im}+\delta\tilde  R_{im}$ in a time interval $\tilde T$ to 
$\tilde T +\delta\tilde T $ by, 
\be
\gamma(\tilde R_{im},\tilde T)\delta \tilde R_{im}\delta\tilde T.
\ee

The size of the small bubble on impact is determined by the initial distance
between the large bubble and the small bubble nucleation site. Thus bubbles with 
impact radius between  $\tilde R_{im}$ and $\tilde R_{im}+\delta\tilde  R_{im}$
were nucleated in a region of width $\delta\tilde  R_{im}$. The spread in nucleation 
time is ismilarly the spread in arrival time. As the nucleation probability per unit
space-time volume  is constant, we have,
\be
\gamma(\tilde R_{im},\tilde T)\sim \gamma(\tilde T)\sim {\rm const.}
\ee
We are interested in the collision rate observed by the bubble dwellers.
For the old bubble we have the usual matter dominated evolution of the scale factor,
$\tilde A_0\sim\tilde\tau^{2/3}$, so the  coordinate time and bubble proper time are related by, 
\be
{d\tilde T\over d\tilde\tau}\propto \tilde\tau^{-2/3}
\ee
In terms of the old bubble proper time, $\tilde\tau$, the collsion rate is,
\be
\gamma(\tilde R_{im},\tilde \tau)\propto  \tilde\tau^{-2/3}.
\ee
If the  scale factor of the smaller bubble at impact is  $a_{im}$, we have, 
$a_{im} \sim \tilde R_{im}/H\tilde A_0$, leading to,
\be
\gamma(a_{im},\tilde \tau)\propto \sim {\rm const}.
\ee
From the point of view of an observer at rest with respect to the centre of the
old bubble, the slowly moving old bubble wall is bombarded a constant flux of
smaller bubbles which are expanding at speeds of order unity.

We assume that the collision of another, smaller  bubble with our own manifests
itself as an injection of energy into some region. 
If the energy injected into the old bubble is proportional to $a_{im}^q$, the
energy spectrum of impacts is given by,
\be
\gamma(E,\tilde \tau)\propto  E^{-1+1/q}.
\ee 
The spectral index depends on $q$, but is always greater than -1.

According to the bubble dwellers, there is a constant rate of collisions with 
young bubbles. The smallest bubbles give the largest impact rate in a given energy interval
and thus provide the most likely experimental signature of such a model. At the opposite
end of the scale we have bubbles with ages upto fractions of a megayear ( as measured
in their own proper times) and sizes upto $10^{-3}$ of our own bubble Universe. 
Collisions with such bubbles would be catastrophic.

\section{Conclusions}

We have considered a brane-world scenario in which our brane is just one of many
bubbles nucleated in an AdS$_5$ bulk. The late time evolution of the scale factor
is Friedmannian. A signature of this particular scenario is the random energy 
injections arising from bubble-bubble collisions. In the $k=1$ case, the rate at 
which small bubbles collide with our own is estimated to vary approximately as 
$E^p$, where E is the energy contained in the impacting bubble and $p>-1$.

\section{Acknowledgements}
The author would like to thank T.J.Hollowood and S.C.Davis for useful discussions.
 
\def\Journal#1#2#3#4{{#1} {\bf #2}, #3 (#4)}
\def\NPB{Nucl.\ Phys.\ B}
\def\PLB{Phys.\ Lett.\  B}
\def\PRL{Phys.\ Rev.\ Lett.\ }
\def\PRD{Phys.\ Rev.\ D}


\begin{thebibliography}{99}
\bibitem{RSa}
L. Randall and R. Sundrum, Phys.Rev.Lett. {\bf 83},  4690 (1999)
\bibitem{RSb}
L. Randall and R. Sundrum,  Phys.Rev.Lett. {\bf 83}, 3370 (1999)
\bibitem{PK}
P. Kraus, JHEP {\bf 9912}, 011 (1999)
\bibitem{DI}
D.Ida, JHEP {\bf 0009}, 014  (2000)
\bibitem{DD}
N. Deruelle and T. Dolezel, gr-qc/0004021
\bibitem{MSM}
S.Mukohyama, T. Shiromizu and K. Maeda, Phys.Rev.D {\bf 62}, 024028 (2000) 
\bibitem{NOZ}
S. Nojiri, S.D. Odintsov and S. Zerbini,  hep-th/0006115
\bibitem{KS}
K.Koyama and J. Soda, Phys.Lett.B {\bf 483} 432 (2000)
\bibitem{GS}
J.Garriga and M. Sasaki,  Phys.Rev.D {\bf 62} 04352  (2000)
\bibitem{CH}
H. Collins and B. Holdom,  hep-th/0006158
\bibitem{Is}
W. Israel, {\it Nuovo Cimento} {\bf 44B} 1 (1966)
\bibitem{ANO}
 L. Anchordoqui, C. Nunez and K. Olsen,   hep-th/0007064

\end{thebibliography}
\end{document}